\documentclass[12pt,a4paper]{article}
\usepackage[T2A]{fontenc}
\usepackage[cp1251]{inputenc}
\usepackage[english]{babel}
\usepackage{amssymb}
\usepackage{amsmath}
\usepackage{graphicx}
\usepackage[small,sc,center]{titlesec}
\usepackage{indentfirst}
\usepackage[labelsep=period]{caption}
\usepackage{caption}
\newcommand{\msun}{\,$M_{\odot}$}

\newcommand{\ergcms}{\,erg\,cm$^{-3}$\,s$^{-1}$}
\newcommand{\ergs}{\,erg\,s$^{-1}$}
\newcommand{\gcmq}{\,g\,cm$^{-3}$}

\newcommand{\kms}{\,km\,s$^{-1}$}

\newcommand{\cmq}{\,cm$^{-3}$}
\newcommand{\cmsqg}{\,cm$^2$\,g$^{-1}$}

\newcommand{\feii}{Fe\,{\sc ii}}
\newcommand{\oi}{O\,{\sc i}}
\newcommand{\oiii}{O\,{\sc iii}}
\newcommand{\oii}{O\,{\sc ii}}

\newcommand{\caii}{Ca\,{\sc ii}}
\begin{document}
	
\begin{center}
\textbf{\large  PISN~2018ibb: radioactive emission of [\oiii] lines}
	\vskip 5mm
\copyright\quad
2024 \quad N. N. Chugai\footnote{email: nchugai@inasan.ru}\\
\textit{ Institute of Astronomy, Russian Academy of Sciences, Moscow} \\
Submitted 11.11.2024 
\end{center}

{\em keywords:\/} stars --- supernovae; stars --- supermassive stars; stars --- nucleosynthesis

\noindent
{\em PACS codes:\/} 

\clearpage
 
 \begin{abstract} 
 Supernova 2018ibb of the PISN category, related to the dynamical instability of oxygen core in a supermassive star induced by pair-creation shows at the nebular stage strong [\oiii] emission lines of an uncertain origin. I propose a simple model that 
 demonstrates a possibility of [\oiii] lines emission from the supernova oxygen matter ionized and heated by the $^{56}$Co radioactive decay.
The reason is pinpointed by which the [\oiii] line luminosity among supernovae of PISN category can vary in  a broad range. 
\end{abstract}

\section{Introduction}

Superluminous supernova (SLSN) SN~2018ibb (Schulze et al. 2024) is the very likely case of 
   pair-instability supernovae (PISN) (Woosley et al. 2002, Barkat et al. 1967)
The light curve in the range of about 1000 days indicates that SN~2018ibb is caused by 
  the supermassive star 
  explosion with the energy of $\gtrsim10^{52}$ erg and the ejection of enormous amount 
   (25--44\msun) of synthesised $^{56}$N  (Schulze et al. 2024).
Large values of energy and $^{56}$Ni mass suggest the presupernova initial mass of 
140--260\msun\ and the explosion is a single event without preliminary pulsations caused by the pair-instability (Woosley et al. 2002). 
This picture leaves no place for any massive circumstellar (CS) shell ejected by a powerful pulsation before the PISN explosion, so one does not expect strong effects of the  CS interaction, unlike SLSN~2006gy with massive ($\sim 5$\msun) CS shell that is a feature   
 of the pulsational pair-instability supernova (PPISN) (Woosley et al. 2007).
 
 In this context one cannot help noticing 
 strong emission lines of [\oiii]\,5007, 4959\,\AA\ and 4363\,\AA\ with the luminosity of $\sim10^{41}$\ergs\ (Schulze et al. 2024).
Among known SLSNe [\oiii]\,5007, 4959\,\AA\ emission apart from SN~2018ibb has been 
  detected only in LSQ14an and PS1-14bj (Lunnan et al. 2016).
It has been suggested that [\oiii] emission of SN~2018ibb could originate from 
 the supernova interaction with a dense CSM (Schulze et al. 2024).
This conjecture, however, does not fully agree with the observed profile  of the 
[\oiii] doublet (Chugai 2024).
The alternative radioactive mechanism of the [\oiii] emission also faces a problem, since 
  the synthetic spectrum based on the explosion model of the He130 helium core 
  (Heger \& Woosley 2002) does not show 
  noticeable [\oiii] 5007\,\AA\ emission (Schulze et al. 2024, Kozyreva et al. 2024).
 
Given uncertainty of the [\oiii] emission origin, it is premature 
  to abandon the radiactive mechanism, because it might well be that the real supernova envelope 
   of SN~2018ibb differs significantly from the theoretical model based on the one-dimensional explosion of the He130 helium core.
First, SN~2018ibb originates from the explosion of the oxygen, not helium core (Chugai 2024).
  Second, the explosion may bring about three-dimensional matter distribution in ejecta; and third, the nickel bubble caused by the decay of the large amount of $^{56}$Ni additionally modifies ejecta density distribution (Kozyreva et al. 2017).      

The above said necessitates to study the radioactive mechanism of 
  [\oiii] emission based on a simple flexible model of SN~2018ibb that 
   takes into account major physics of the radioactive energy deposition, including oxygen ionization and excitation.
 This study is the goal of the present communication that includes the description of the 
 simplified model of SN~2018ibb and the computation method for the oxygen ionization (Section 2),
 the calculation of oxygen line intensities and the determination of principal parameters 
 of line-emitting oxygen (Section 3). 
 Results are discussed in Section 4, where I present also an explanation of the significant 
 variation of the [\oiii] lines intensity among possible supernovae of the PISN category.  

\section{Oxygen ionization}

The supernova envelope is represented by a homogeneous (on average) sphere with the free  
   expansion kinematics $v = r/t$. 
 The model is essentially one-zone with the average density determined by the mass and   
    kinetic  energy. 
 The referential SN~2018ibb model is mod60 (Chugai 2024) with the ejecta mass of $M = 60$\msun, 
  energy $E = 1.2\times10^{52}$\,erg, and $^{56}$Ni mass $M_{ni} = 30$\msun.
It should be emphasised that this model is also an approximation based  on 
  the Arnett (1980) description.
Principal parameters of our model are consistent qualitatively with  
  the  explosive burning of 30\msun\ that releases 
  $3\times10^{52}$\,erg  for the  specific energy $q(^{16}O) =5\times10^{17}$\,erg\,g$^{-1}$
  (Woosley et al. 2002).  
With the subtracted kinetic energy the remaining $1.8\times10^{52} $\,erg  is spent 
  on the bound energy, which at the maximum compression significantly exceeds  
  the bound energy of 60\msun\ oxygen core prior to the loss of stabiliy.

For the indicated mass and energy the boundary velocity is  $v_0 = (10E/3M)^{1/2} = 5770$\kms, the average density 
 is $\rho = 3.6\times10^{-15}(t/400\mbox{d})^{-3}$\gcmq, and the oxygen number density 
 $n = 1.3\times10^8(t/400\mbox{d})^{-3}$\cmq.
We consider three cases of the oxygen mass:   $M_O = 15$\msun, 10\msun, and 5\msun.
The oxygen is presumably mixed with $^{56}$Ni macroscopically, 
 viz., any volume of the oxygen material consists only of the oxygen.
This requirement suggests that the relative distance $\delta = \Delta v/v_0$ between 
  fragments of $^{56}$Ni matter and the oxygen should permit gamma-quanta from the $^{56}$Co 
   decay to penetrate in the oxygen avoiding the noticeable absorption at the considered stage $t \approx 400$ days.
In other words, the following condition should be met: $\Delta \tau_{\gamma} = \delta k_{\gamma}\rho v_0t \approx 2.2\delta < 1$, where the absorption coefficient for gamma-rays of $^{56}$Co decay is $k_{\gamma} = 0.03$\cmsqg\ (Sutherland \& Wheeler 1984).
The requirement $\delta < 0.5$ is fulfilled even for the moderate macroscopic mixing.

The approximation of a homogeneous density is an idealization.
In fact the density distibution can be essentially three-dimensional due to 
 effects of the explosion and nickel bubble, so we admit that the oxygen density 
  ($\rho_O$) can differ from the average density ($\rho$) by a factor $\chi$ of the density contrast $\rho_O = \chi \rho$. 
With the oxygen mass $M_O < M$ the oxygen filling factor is $f = V_O/V =  
  (M_O/M)\chi^{-1}$. 
Note that for $\chi = 1$ the oxygen filling factor $f = M_O/M < 1$, which clarifies   
   the use of $\chi$ instead of the filling fsctor $f$. 

At the age $t = t_{max} + 276 = 381$ days after the explosion [+276 days mean 276 days  after the bolometric maximum $t_{max}$ (Schulze et al. 2024), while 381 days is the time lapse after the explosion] 
  the SN~2018ibb bolometric luminosity is $L = 1.2\times10^{43}$\ergs\ (Schulze et al. 2024) that is reproduced in the model mod60 (Chugai 2024). The power deposited in the oxygen matter by gamma-quanta of 
 $^{56}$Co decay is $L_d(\mbox{O}) = (M_O/M)L$.
The deposiion in the unit of volume of the oxygen matter is, therefore,  $\epsilon_d =  L_d(\mbox{O})/V_O) = (L/V)\chi = 3.6\times10^{-7}\chi$ [\ergcms].
 
The ionization rate of the oxygen $k$-ion with the ionization potential $I_k$ due to  
  ionization losses of Compton electrons is $y_k\epsilon_d/w_k$, where $y_k$ is the 
  $k$-ion fraction and $w_k$ is the average work spent on the single ionization of the $k$-ion.
The experimental data for the neutral oxygen with the ionization potential $I_1$  
  suggest $w_1 = 2.3I_1$ (Ahlen 1980); the similar relation $w_k = 2.3I_k$ is used below for ions 
  $k = 2$ and $k = 3$ since the ionization by fast electrons with the energy 
  $\mathcal{E} \gg I_k$ removes most loosely bound  external electron.
The radiative recombination rate is $R_k = \alpha_k n_ey_{k+1} n$, where $n_e$ is 
 the electron number density calculated recursively in the process of solution of 
 ionization balance equations, $n$ is the oxygen number density, $\alpha_k$ is the 
 recombination coefficient for $k$-ion taken from (Tarter  1971).
 
In conditions under consideration the oxygen ioniztion occurs in the steady-state regime.
Indeed, for minimum ionization degree on day 400 $x_e = n_e/n = 0.1$ (in reality, $x_e \gtrsim 0.5$) 
  the maximum recombination time is $1/(\alpha_1n_e) \sim 4\times10^5$\,s that is 
  far below both the $^{56}$Co lifetime ($\sim 10^7$\,s) and the expansion time.
In the steady-state regime equations of the ionization balance read $G_k = R_k$.
The system of ionization balance equations includes four ionization stages of oxygen. 
As an example, on day 400 for the model with the oxygen mass of 15\msun\ and fiducial parameters $T_e = 9000$\,K and $\chi = 3$ the ionization fractions are  
 ($y_1, y_2, y_3, y_4$) = (0.439, 0.52, 0.0397, 0.0013) with the resulting ionization 
 degree $x_e = 0.602$.

Note that we ignore additional ionization of \oi\ by ultraviolet resonance lines of 
 the multiplet uv1 \oii\ (834\,\AA) emitted due to the excitation of \oii\ by fast electrons.
As a result, our model underestimates a bit the oxygen ionization degree.
 
%
\begin{figure}
	\centering
	\includegraphics[trim=0 100 0 270,width=\columnwidth]{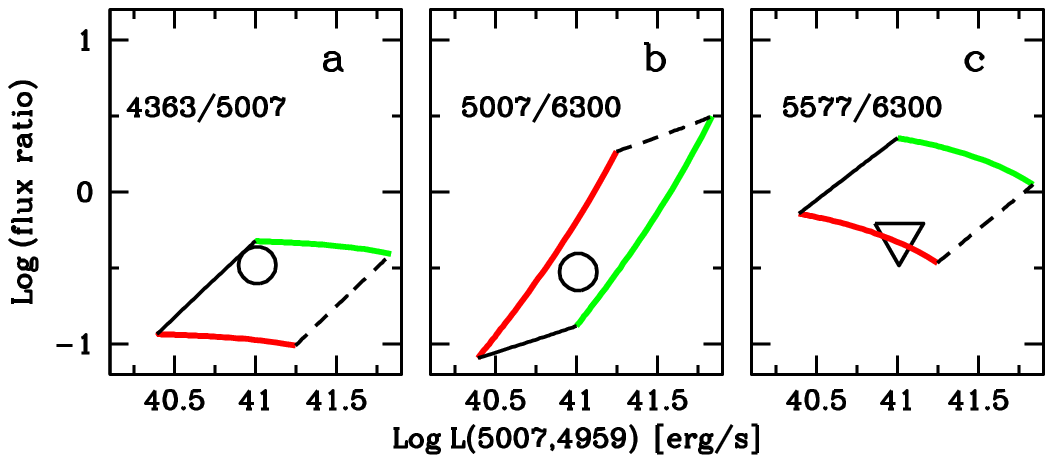}
	\caption{
		Diagnostic diagram "flux ratio vs.~[\oiii] line luminosity" for the 
		model with 15\msun\ of oxygen. 
		{\em Left} ({\bf a}) is the diagram for the flux ratio 	[\oiii]\,4359\AA/5007,4959\AA; 
			{\em center} ({\bf b}) for the flux ratio 
			[\oiii]\,5007,4959\AA/[\oi]\,6300,6364\AA; 
		 {\em right} ({\bf с}) for the ratio  
		 [\oi]\,5577\AA/[\oi]\,6300,6364\AA.
		{\em Circles} show observational values, {\em triangle} is for the upper limit of 5577\AA/[\oi]\,6300,6364\AA\ flux ratio.
		{\em Colored } lines show the dependence on the  $\chi$ parameter for the 
		fixed temperature (red is for 7000\,K and green is for 10000\,K).
		{\em Black solid} line shows the dependence on the temperature for fixed 
		density contrast $\chi = 5$,  while {\em dashed} line is for $\chi = 0.5$.
			}
	\label{fig:flux1}
\end{figure}

\section{Oxygen line intensities}

The ion fractions and electron number density found by the solution of ionization balance equations 
  for given values of $T_e$ and $\chi$ are used for the calculation of oxygen luminosity in lines of [\oi], [\oii], and [\oiii].
Level populations are obtained from balance equations in three-level approximation 
(level corresponds to term).  
Rates of spontaneous transitions are from NIST database, collisional   
  strengths are from data compiled in the book (Osterbrock \& Ferland 2006), 
  while for \oi\ collisional strength are from (Zatsarinni \& Tayal 2003).
The solution of balance equations with radiative and collisional transitions provides  
  us with the line power density ($\epsilon = 4\pi j$, where $j$ is the emissivity) for given values of $T_e$ and $\chi$.
  The line luminosity is the production of $\epsilon$ and the oxygen volume 
$L = \epsilon V(M_O/M)\chi^{-1}$.

Results for the oxygen mass of 15\msun\ are displayed (Figure 
 \ref{fig:flux1}) as diagrams "line flux ratio vs. [\oiii] 5007,4959\,\AA\ luminosity".
The observational luminosity in [\oiii] double is lg\,$L = 41.01$ [\ergs], whereas 
  the flux ratio $F(4359)/F(5007б4959) = 0.3\pm0.3$. 
Both values are reproduced for $T_e \approx 9000$\,K and $\chi \approx 3$ 
(Figure \ref{fig:flux1},a).
Note that the  $F(4359)/F(5007,4959)$ ratio is a sensitive indicator of the electron temperature.

The luminosity of [\oiii] doublet and the flux ratio of [\oiii]/[\oi] 
 (Figure \ref{fig:flux1},b) are formally reproduced in the model $M_O = 15$\msun\ 
 for $T_e \approx 8800$\,K and $\chi \approx 3$.  
In fact, the profile of [\oi] is significantly broader compared to [\oiii] emission 
  (Schulze et al. 2024) indicating a different radial distribution of sources for these emissions, which is not described by our one-zone model.
  
The auroral-to-nebular ratio [\oi] 5577\AA/6300,6364\AA\ (Figure \ref{fig:flux1},c) 
is an indicator of the tmperature and electron number density in the [\oi]
 emission zone.
 Unfortunately, at the stage +276 days the 5577\,\AA\ is not seen because it falls on the red slope of broad emission band of \feii\ lines; the auroral line becomes visible 
  only at the late stage $t \geq t_{max} + 377$ days (cf. Schulze et al. 2024).
The upper limit of the 5577\AA/6300,6364\AA\ ratio indicates the electron temperature 
  $\lesssim 7000$\,K in the [\oi] line-emitting zone (Figure \ref{fig:flux1},c). 
We do not show comparison of the computed [\oii]\,7325\,\AA\ luminosity with the observational one 
  because this line is blended with the strong emission of [\caii] 7300\,\AA. 
Yet it is noteworthy that the calculated  [\oii]\,7325\,\AA\ luminosity does not contradict 
 to the significant contribution of this line to the 7300\,\AA\ emission.

The model with the oxygen mass of 10\msun\ reproduces the [\oiii] doublet luminosity and 
  the [\oiii] 4359\AA/5007,4959\AA\ ratio for parameter values 
$T_e \approx 9000$\,K and $\chi \approx 2$ comparable to those of the model with 15\msun\ of 
  oxygen.
We therefore do not show figures for this case.
For the model with the oxygen mass of 5\msun\ the luminosity of the [\oiii] 
doublet and [\oiii] 4359\AA/5007,4959\AA\ ratio are reproduced for somewhat different 
 parameters: $T_e \approx 9200$\,K and $\chi \approx 1$  (Figure \ref{fig:flux1},a). 
The large model ratio of [\oiii]/[\oi] lines (Figure \ref{fig:flux1},b)
for these parameters reflects the strong \oi\ ionization due to the relatively low density.

Input parameters and inferred values are summarized in Table 1.
It shows, in order, the  oxygen mass, temperature, density contrast, ionization degree, power  
 deposition ro oxygen material, total luminosity of oxygen lines, deposition fraction estimated  from the energy thermal balnce $\eta_h = L_{em}(\mbox{O})/L_d(\mbox{O})$, and finally, the theoretical estimate using derived ionization degree and the  relation  $\eta_{h,num}(x_e)$ 
 for the oxygen found earlier via the numerical solution of Spenser-Fano equation (Kozma \& Fransson 1992).
In the model with $M_O = 15$\msun\ $\eta_{h,num}$ value is equal to $\eta_h$ (Table 1), 
for the model with $M_O = 10$\msun\ these parameters coincide within 20\%. 
In the model with $M_O = 5$\msun\ the difference mounts to a factor of 1.5. 
The latter indicates that the model with the large (10--15\msun) oxygen mass 
is preferred. 
The coincidence of $\eta_h$ and  $\eta_{h,num}$ for sensible values of model parameters 
  supports the radioactive mechanism of [\oiii] emission from SN~2018ibb ejecta.
%
\begin{figure}
	\centering
	\includegraphics[trim=0 110 0 250,width=\columnwidth]{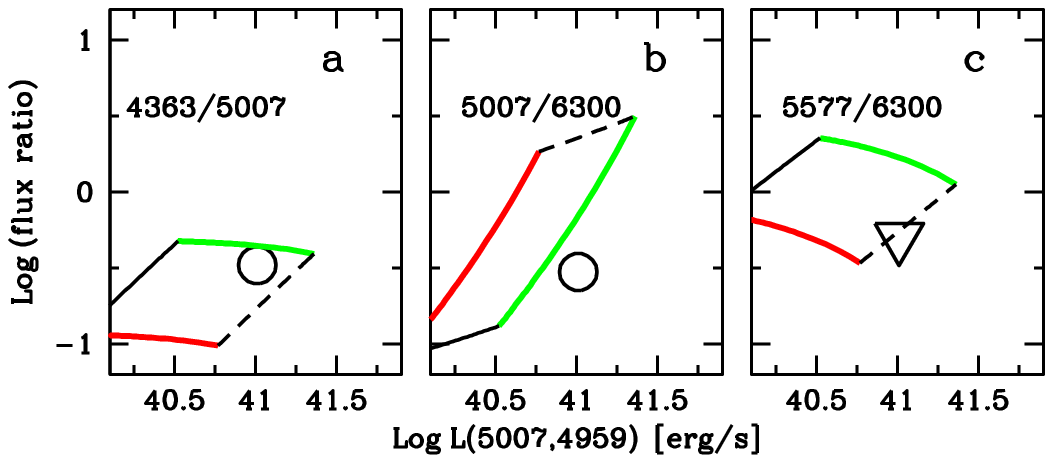}
	\caption{
			The same as Figure \ref{fig:flux1}, but with the oxygen mass of 5\msun.		
	}
	\label{fig:flux2}
\end{figure}
 %
 %
\begin{table}
	\vspace{6mm}
	\centering
	{{\bf Table 1.} Model parameters and oxygen luminosity}\\
	\bigskip	 
	\begin{tabular}{|p{1.5cm}ccccccc|} 
		\hline	
	  $M_O/M_{\odot}$ & $T_e$  & $\chi$ & $x_e$ & $L_d(\mbox{O})$  & $L_{em}(\mbox{O})$  &  $\eta_h$ & $\eta_{h,num}$ \\
	\hline		
		15    & 9000  &  3.3  & 0.59  &  3(42)$^{\bigstar}$  &  2.3(42)   &  0.77 & 0.75\\
		10     & 9000  & 2  & 0.69  &  2(42)  &   1.29(42)  &  0.63 &  0.77 \\
	   5     & 9200   & 1  & 0.86  &  1(42)  &   5.2(41)  &  0.51 &  0.79 \\	
		\hline
	\end{tabular}
	\medskip
	
	\begin{tabular}{l}
		$^{\bigstar}$ Luminosity in \ergs\ (in parentheses is the power of 10)\\	
	\end{tabular}
\end{table}
%

\section{Discussion}

The goal of the paper has been a question, whether the strong forbidden emission of double-ionized oxygen in SN~2018ibb could be explained in the framework of the radioactive mechanism.
I find that one-zone model with realistic values of supernova energy, mass and 
 amount of $^{56}$Ni is able to describe the observed luminosities of [\oiii] lines, provided the macroscopic mixing between $^{56}$Ni and oxygen, and thus to confirm the radioactive mechanism of these emissions.
 
The valuable feature of our one-zone model is the minimum set of free parameters, which are $T_e$ and $\chi$.
Yet one-zone model disadvanage is that it cannot account for some important issues.
First, the profile of the [\oi] doublet is broader compared to [\oiii] doublet 
 (cf. Schulze et al.2024), which suggests somewhat different line-emitting zones. 
[\oi] doublet forms predominantly in outer high-velocity supernova layers, whereas 
 [\oiii] lines form in the inner zone. 
To describe this picture at least two-zone model is needed. 
Second, the one-zone model with the $T_e \approx 9000$\,K predicts rather strong 
[\oi]\,5577\,\AA\ line at the stage +276 days, which is absent in the spectrum and 
 appeares only after +360 days (cf. Schulze et al. 2024).
The relatively low flux ratio  5577\AA/6300,6364\AA\ emphasises that line-emitting zones 
of [\oi] and [\oiii] do not coincide and  the temperature in [\oi] line-emitting zone is  
lower, $T_e \lesssim 7000$\,K.
Both aforementioned points require two-zone model, in which internal "hot" oxygen with 
  the temperature of 9000\,K emits [\oiii] lines, whereas external "cold" oxygen contributes 
  primarily in [\oi] line emission.
 Noteworthy, the disparity between emission of [\oiii] and [\oi] nave been noticed in the model
 of LSQ14an (SLSN) spectrum: the model with low-mass ejecta reproduces [\oiii] but significantly 
   underproduces [\oi] emission (Jerkstrand et al. 2017).
   
 Surprisingly, at first glance, that the complicated model of the radioactive emission of oxygen   
  (Jerkstrand et al. 2017) is not able to reproduce the [\oiii] luminosity of SN~2018ibb despite being based on the hydrodynamic PISN model. 
A possible reason is that the one-dimensional PISN model does not produce the significant 
  macroscopic mixing of the $^{56}$Ni in the unburned oxygen.
Meanwhile such a mixing is a crucial ingredient that significantly increases the gamma-quanta  deposition into the oxygen material compared to the unmixed case.   

It is interesting to understand conditions favoring the emergence of the strong [\oiii] emissions in 
 supernovae of PISN category.
Let us consider the ionization balance between \oii\ and \oiii\ with the relative fractions 
$y_2$, $y_3$ and the electron number density $n_e \approx y_2n$
\begin{equation}
\frac{\epsilon_d y_2}{w_2} = \alpha_2 y_2y_3n^2\,.
\label{eq:ion1}
\end{equation} 
Taking into account volumetric power $\epsilon_d \propto M_{ni}/V$, volume 
 $V \propto E^{3/2}/M^{3/2}$, and the result that the mass of synthesised 
 $^{56}$Ni for model PISN with helium cores 100-130\msun\ relates to the 
  explosion energy approximately as $M_{ni} \propto E^3$ (Kasen et al. 2011), 
   we obtain the relation  
\begin{equation}
	y_3 \propto M_{ni}^{3/2}M^{-7/2}\,,
\label{eq:ion2}	
\end{equation}  
The found relation says that a higher fraction of \oiii\ and therefore the larger 
 flux ratio of lines 5007,4959\AA/6300,6364\AA\ is expected in PISN supernovae with  the larger $^{56}$Ni mass and the lower ejecta mass. 
The explosion of a bare oxyygen core with a large  $^{56}$Ni mass in the case of 
SN~2018ibb creates favorable conditions for the emergence of the
 strong [\oiii]\,5007,\,4959\AA\ emission.

The relation (\ref{eq:ion2}) permits one to understand, why among PISNe the relative 
 ratio of [\oiii]/[\oi] emissions can vary in a broad range.
Let us imagine that in some supernova of PISN category the ejecta mass, due to the helium shell, 
is twice as large, while the $^{56}$Ni mass is twice as low compared to  SN~2018ibb.
The fraction of \oiii\ ion then will be $\approx 30$ times lower and 
by the same factor the flux ratio 5007,4959\AA/6300,6364\AA\  will be lower compared to SN~2018ibb.

\section{Conclusion}

The analysis of the oxygen ionization and excitation based on a simple model of the 
  SN~2018ibb envelope with realistic values of ejecta mass, energy and amount of $^{56}$Ni 
  lead us to conclude that observed [\oiii]
 emission lines can originate from the deposition of the radioactive energy of the $^{56}$Co  decay into oxygen matter given the macroscopic mixing between $^{56}$Co and  oxygen.
It is shown that a significant range of flux ratio of [\oiii]/[\oi] lines 
 5007,4959\AA/6300\AA\ 
  among possible supernovae of PISN category is due to the difference of ejecta mass and synthesised $^{56}$Ni mass.
The increase of \oiii\ fraction with the larger mass of $^{56}$Ni and lower ejecta mass   
 explains, why SN~2018ibb has strong [\oiii] emission lines.

\clearpage

\section{References} 

\noindent
\medskip
Ahlen S. P., Rev. Mod. Phys. {\bf 52}, 121 (1980)\\
\medskip  
Barkat Z., Rakavy G., Sack N., Phys. Rev. Letters {\bf 18}, 379 (1967)\\
\medskip 
Chugai N. N., ArXiv e-prints [arXiv:2410.17580] (2024)\\  
\medskip 
Heger A., Woosley S. E., Astrophys. J. 567, 532 (2002)\\
\medskip
Jerkstrand A., Smartt S. J., Inserra C.  et al.,  Astrophys. J. {\bf 835}, 13 (2017)\\
\medskip 
Kasen D., Woosley S. E., Heger A., Astrophys. J. {\bf 734}, 102 (2011)\\
\medskip 
Kozma C., Fransson C.,  Astrophys. J. {\bf 390}, 602 (1992)\\
\medskip
Kozyreva A., Shingles L., Baklanov P. et al., Astron. Astrophys. 689, A60 (2024)\\
\medskip
Kozyreva A., Gilmer M., Hirschi R. et al., Mon. Not. R. Astron. Soc. {\bf 464}, 2854 (2017)\\
\medskip
Lunnan R., Chornock R., Berger E. et al., Astrophys. J.831, 144 (2016)\\
\medskip
Osterbrock D. E., Ferland G. J. {\em Astrophysics of gaseous nebulae and active galactic nuclei}, (Sausalito, CA: University Science Books) (2006) \\
\medskip
Schulze S., Fransson C., Kozyreva A., et. al., Astron. Astrophys. {\bf 683}, A223 (2024)\\
\medskip
Sutherland P. G., Wheeler J. C., Astrophys. J. {\bf 280}, 282 (1984)\\
\medskip
Tarter S. B., Astrophys. J. {\bf 168}, 313 (1971)\\
\medskip
Woosley S. E., Blinnikov S., Heger A., Nature, {\bf 450}, 390 (2007)\\
\medskip
Woosley S. E., Heger A., Weaver T. A., Review Mod. Phys. {\bf 74}, 1015 (2002)\\
\medskip
Zatsarinni O., Tayal S. S., Astrophys. J. Supl. Ser. {\bf 148}, 375 (2003)\\

\end{document}